\documentclass[twocolumn]{emulateapj}

\newcommand{\FP}{Fabry-Perot}
\newcommand{\ghafas}{{\sc GH$\alpha$FaS}}

\newcommand{\Ha}{H$\alpha$}

\newcommand{\HI}{{\sc H$\,$i}}
\newcommand{\HII}{{\sc H$\,$ii}}
\newcommand{\CO}{{\sc CO}}

\newcommand{\Vlos}{$V_\mathrm{los}$}
\newcommand{\Vsys}{$V_\mathrm{sys}$}
\newcommand{\Vrot}{$V_\mathrm{rot}$}
\newcommand{\Vrad}{$V_\mathrm{rad}$}
\def\mpcyr{$M_\sun\mbox{ pc}^{-2}\mbox{ yr}^{-1}$}
\def\kms{$\mbox{km s}^{-1}$}
\def\kmskpc{$\mbox{km s}^{-1}\mbox{ kpc}^{-1}$}

\def\deg{^\circ}

\shorttitle{Spiral inflow feeding the nuclear starburst in M\,83}
\shortauthors{Fathi et al.}

\slugcomment{Accepted for publication in ApJ Letters}
\begin{document}

   \title{Spiral inflow feeding the nuclear starburst in M\,83, observed in \Ha\ emission with the \ghafas\ Fabry-Perot interferometer}

   \author{Kambiz Fathi\altaffilmark{1,2},
   	  John E. Beckman\altaffilmark{1,3},
	  Andreas A. Lundgren\altaffilmark{4},
          Claude Carignan\altaffilmark{5},
          Olivier Hernandez\altaffilmark{5}, 
	  Philippe Amram\altaffilmark{6},
	  Philippe Balard\altaffilmark{6},
	  Jacques Boulesteix\altaffilmark{6},
	  Jean-Luc Gach\altaffilmark{6},
	  Johan H. Knapen\altaffilmark{1},
	  Monica Rela\~no\altaffilmark{7}}
\altaffiltext{1}{Instituto de Astrof\'\i sica de Canarias, 
		 C/ V\'\i a L\'actea s/n, 38200 La Laguna, Tenerife, Spain. 
		 Email: fathi@iac.e, jeb@iac.es, jhk@iac.es}
\altaffiltext{2}{Stockholm Observatory, AlbaNova 
		 University Center, 106 91 Stockholm, Sweden.}
\altaffiltext{3}{Consejo Superior de Investigaciones Cient\'\i ficas, Spain.}
\altaffiltext{4}{ESO, Casilla 19001, Santiago 19, Chile. 
		 Email: alundgre@eso.org}
\altaffiltext{5}{LAE, D\'epartement de Physique, 
		 Universit\'e de Montr\'eal, C.P. 6128 succ. 
		 center ville, Montr\'eal, QC, Canada H3C 3J7. 
		 Email: olivier@astro.umontreal.ca, claude.carignan@umontreal.ca}
\altaffiltext{6}{LAM, 2 Place Le Verrier, 13248 Marseille Cedex 04 France. 
		 Email: philippe.amram@oamp.fr, philippe.balard@oamp.fr, 
		 jacques.boulesteix@oamp.fr, jean-luc.gach@oamp.fr}
\altaffiltext{7}{Departamento de F\'{\i}sica Te\'oretica y del Cosmos, 
		 Universidad de Granada, Spain. 
		 Email: mrelano@ugr.es}

\begin{abstract}
We present observations of the nearby barred starburst galaxy, M\,83 (NGC\,5236), with the new Fabry-Perot interferometer \ghafas\ mounted on the 4.2 meter William Herschel Telescope on La Palma. The unprecedented high resolution observations, of 16 pc/FWHM, of the \Ha-emitting gas cover the central two kpc of the galaxy. The velocity field displays the dominant disk rotation with signatures of gas inflow from kpc scales down to the nuclear regions. At the inner Inner Lindblad Resonance radius of the main bar and centerd at the dynamical center of the main galaxy disk, a nuclear $5.5 (\pm 0.9) \times 10^8 M_\sun$ rapidly rotating disk with scale length of $60 \pm 20$ pc has formed. The nuclear starburst is found in the vicinity as well as inside this nuclear disk, and our observations confirm that gas spirals in from the outer parts to feed the nuclear starburst, giving rise to several star formation events at different epochs,  within the central 100 pc radius of M\,83.
\end{abstract}


\section{Introduction}
\label{sec:introduction}
The tendency of gas to flow into the potential well at the center of a galaxy is, in physical terms, expected, however the details of the mechanisms which fuel active galactic nuclei and circumnuclear starbursts are not easy to disentangle. This is not due to lack of theoretical insight but because the relevant observations are difficult to acquire. While the Hubble Space Telescope has mapped pc scale morphological features, the corresponding kinematic observations are far harder to come by. The mechanisms which we must test imply the transfer of angular momentum from the inflowing gas to its surroundings. A key scenario was formulated  by Shlosman et al. (1990) showing that non-axisymmetric circumnuclear structures can be the last link in a chain of inflow in large bars. Englmaier \& Shlosman (2000) showed that in galaxies with low central mass concentrations dynamical processes would not favor the formation of central bars, but might instead give rise to nuclear spiral structure in the gas, whose exact parameters depend on the mass distribution and the sound speed.\looseness-2

In the late 1990's, spiral patterns were found around the nuclei of many galaxies, some flocculent, such as M\,51 (Grillmair et al. 1997), others of grand design connected to outer spiral structure (Laine et al. 1999). A large bar can, according to Englmaier \& Shlosman (2000) drive stable spiral morphology in gas well within the inner Lindblad resonance (ILR) radius, right down to the nuclear zone. This inflow could produce a gaseous nuclear bar even without the presence of a stellar bar. An additional twist was given by Beck et al. (2005) who detected magnetic field lines aligned with the strong bars of NGC\,1097 and NGC\,1365, and in the circumnuclear zone of NGC\,1097 found an in-spiralling field of strength 60 mG which they claimed could funnel the gas into the active nucleus (AGN).\looseness-2

Morphological observations show that tightly wound nuclear dust spirals occur mainly in weakly barred galaxies, while strongly barred galaxies may have grand design spirals which do not penetrate down to the nucleus but can end at a circumnuclear starburst ring (e.g., Laine et al. 1999;2001; Peeples \& Martini 2006). However, \HI\ and \CO\  observations (with few exeptions) cannot resolve this kind of structures on the required arcsecond scales. Recently, Zurita et al. (2004) demonstrated spiral inflow, along dust lanes, towards the nuclear regions of the strongly barred galaxy NGC\,1530, and Fathi et al. (2006) quantified the kinematics of  spiral inflow down to 10 pc from the active nucleus in NGC\,1097. In both of these cases no nuclear bar is present, and the velocity vectors spiral in along the opposing arms, as predicted by Englmaier \& Shlosman (2000). \looseness-2

\begin{figure*}
\includegraphics[width=1.0\textwidth]{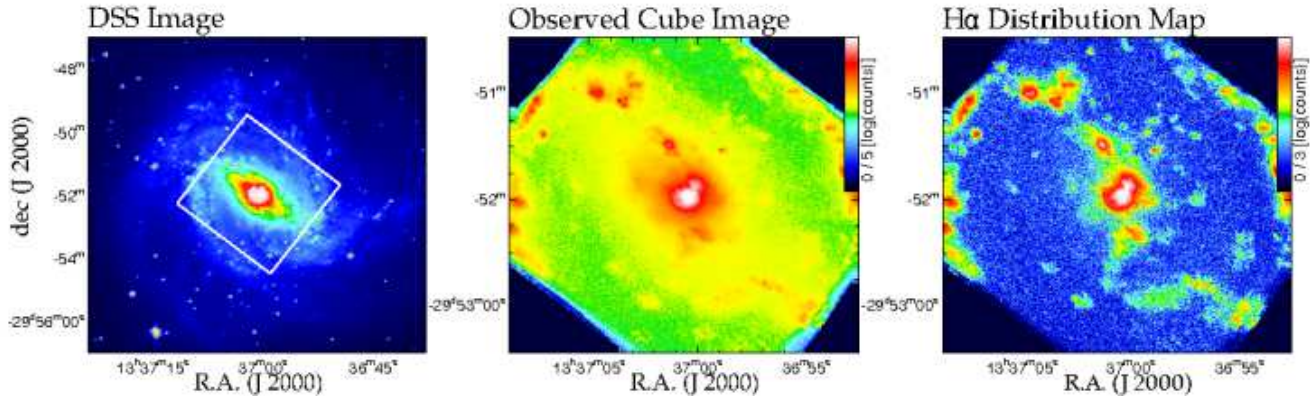}
\caption{Digitized Sky Survey image with the \ghafas\ footprint (left), the data cube image (middle), and the \Ha\ map (right).}
\label{fig:obs}
\end{figure*}

While commissioning the rapid scan Fabry-Perot spectrograph \ghafas\ (Hernandez et al. 2007) on the William Herschel Telescope (WHT), we obtained a data cube in intensity and velocity for the \Ha\ emission from the central two kpc of the nearby grand design spiral galaxy M\,83 (NGC\,5236). We analyze the kinematics the circumnuclear zone of M\,83   to see whether we can detect the gas which feeds the nuclear starburst studied in detail by Harris et al. (2001). Here we give a brief description of the instrument and the data handling , and show that we could  detect interlocking spiral inflow velocities down to 10 pc from the nucleus of this galaxy, then outline some   implications of the measurements.\looseness-2

\begin{table}
  \label{tab:data}
  \begin{center}
  \caption{Galaxy, observing log, and main data quality parameters.}
    \begin{tabular}{ll} \hline \hline              
Classification; $B$-band magnitude (NED)            	& SAB(s)c/Sbrst; 8.2	\\
Total dynamical mass {(L04b)}			& $6.1\times 10^{10} M_\sun$	\\
Distance (Thim et al. 2003)			& 4.5 Mpc			\\
Heliocentric velocity (This Study)		& $520\pm13$ \kms 		\\
\hline
Mean seeing (FWHM)	      	& $0.7\pm0.1$ \arcsec		 		\\
Pixel size			& 0.4 \arcsec/pix				\\
Field-of-view 			& $3.4\arcmin \times 3.4\arcmin$		\\
\hline
Interference order (at \Ha)   	& $765\pm 1$		 \\
Mean Finesse (at \Ha) 	     	& 16.7			 \\
Channel width	     		& 8.16 \kms		 \\
Free spectral range  		& 391.89 \kms		 \\
Number of cycles/channels	& 15/48			 \\
Total exposure time on object	& 60 min		 \\
\hline 
    \end{tabular}
  \end{center}
\end{table}
\section{Observations and Data Reduction}
\label{sec:observations}
We observed M\,83 on July 6$^{th}$ 2007 with the \ghafas\ Fabry-Perot interferometer mounted on the GHRIL platform of the Nasmyth focus of the 4.2m WHT (Hernandez et al. 2007). The observations were done in non-photometric conditions, on average 20\% air humidity, and with seeing variations between 0.6\arcsec and 0.9\arcsec. The instrument specifications and setup together with the main galaxy parameters are presented in table~\ref{tab:data}.

\begin{figure*}
\includegraphics[width=0.99\textwidth]{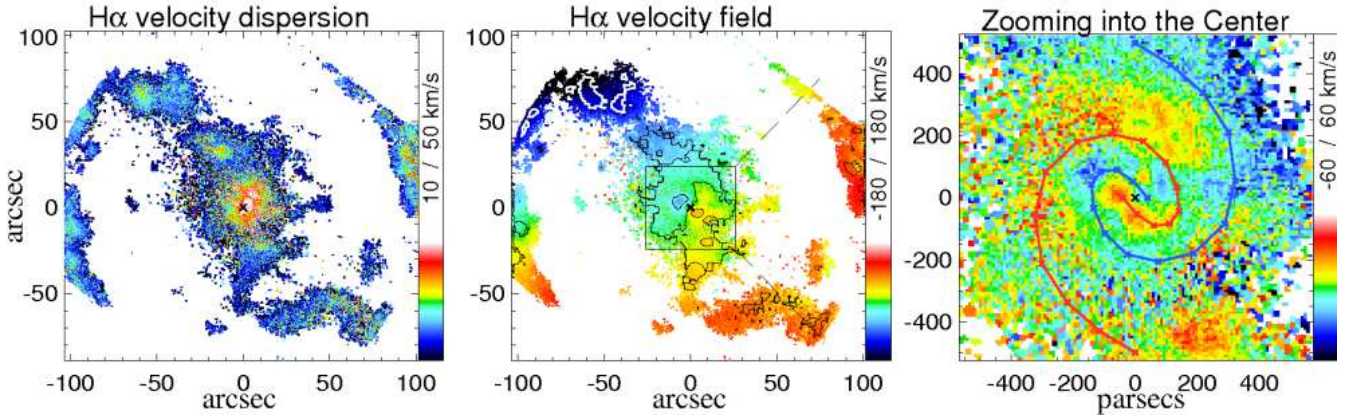}
\caption{\Ha\ kinematic maps displaying pixels with $ A/N > 10$ and the position of the dynamical center with a cross (section~\ref{sec:quantify}). Removing the inner disk (Fig.~\ref{fig:vrot}), the spiral inflow is aligned with the outer velocity field, and can be followed down to the nuclear starburst (right panel). All panels are oriented as Fig.~\ref{fig:obs}, and to guide the eye, the inspiral trajectory is outlined. The contours mark velocity steps of 40 \kms.}
\label{fig:kinematics}
\end{figure*}

The \ghafas\ data acquisition and reduction involves the standard photon counting Fabry-Perot steps which we perform by use of the software package of Daigle et al. (2006) and our own pipeline described in Hernandez et al. (2007). \ghafas\ has a field-of-view (FOV) of $3.4\arcmin \times 3.4\arcmin$, with 0.4\arcsec/pix, corresponding to 9 pc/pix, and 16 pc/FWHM at the galaxy. Since M\,83 is much larger than our FOV, sky lines and sky variations were sampled by getting one cycle every three of a blank field close to the galaxy. The sky cubes are thereafter reduced and subtracted from the galaxy in four different ways i.e., constant sky spectrum, and 1st, 3rd and 4th order polynomials fitted to each channel image (Daigle et al. 2006; Hernandez et al. 2007). Each reduced galaxy cube was treated independently to derive the \Ha\ kinematics by deriving the moment maps with consistency checks by fitting single Gaussian profiles to the spectra (e.g., Fathi et al. 2007). In order to gain more field coverage, we applied the two-dimensional Voronoi tessellation method on the faint regions and cleaned the maps by removing all values which are derived from spectra with amplitude-over-noise ($A/N$) less than 10. Although this seems a strict criterion, we choose to continue our analysis of the velocity field based on spectra which are reliable not only for deriving reliable velocities, but also velocity dispersion. The instrumental dispersion, $\sigma_\mathrm{inst}=2.5$ \kms, measured using a Neon lamp, was subtracted quadratically from the galaxy dispersion (see Fig.~\ref{fig:kinematics}).

\section{Results}
\label{sec:results}
The \ghafas\ observations clearly show the high \Ha\ intensities at the known star-forming regions in M\,83, the positions of which are in good agreement with the well known kpc-size stellar bar in the north-east direction, (Fig.~\ref{fig:obs}), first observed with \FP\ interferometry by Carranza (1968). At the center, where the star formation activity is higher, we observed the highest \Ha\ intensity. The velocity field clearly displays the disk rotation as the dominant kinematic feature, as well as a rapidly rotating component in the inner 20\arcsec\ radius from the nucleus followed by a kinematic position angle twist at its edges (Fig.~\ref{fig:kinematics}). The velocity dispersion map displays several local maxima at the location of the star-forming \HII\ regions, accompanied with a global maximum coinciding with the derived dynamical nucleus of the galaxy.

\subsection{Quantifying the Kinematics}
\label{sec:quantify}
\begin{figure}
\includegraphics[width=0.49\textwidth]{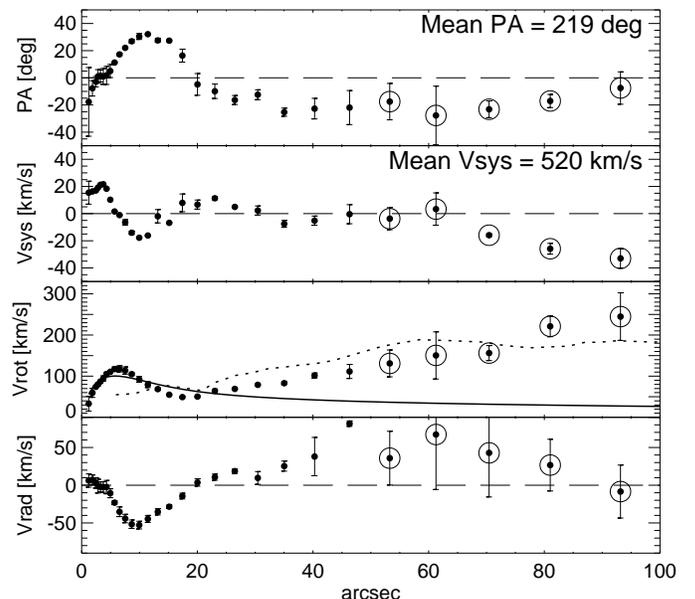}
\caption{The kinematic parameter profiles described in section~\ref{sec:quantify}. Overplotted on the \Vrot\ are the \CO\ rotation curve from {L04b} (dotted curve), and the best fitted inner disk profile (solid curve). Outside a radius of 50\arcsec, uneven sampling of the field leads to unreliable kinematic parameter values.}
\label{fig:vrot}
\end{figure}

To quantify the observed velocity field, we apply the tilted ring decomposition combined with the harmonic decomposition formalism (Schoenmakers et al. 1997;Fathi et al. 2005; and Krajnovi\'c et al. 2006). Accordingly, we divide the velocity field in concentric rings and fit the disk geometry followed by fitting the circular and non-circular motions within each ring simultaneously. We fix the inclination at $24\deg$ (Lundgren et al. 2004b, hereafter {L04b}) and derive the dynamical center, systemic velocity (\Vsys), and position angle (PA) iteratively. Once these parameters have been found, we fix them and for each concentric ring derive the rotational and radial velocity components (\Vrot\ and \Vrad) and up to and including the third-order harmonic terms. For each ring, we fit both the circular and non-circular velocities at the same time (Fig.~\ref{fig:vrot}). To investigate how choice of parameters in the reduction procedure affects the final results, we applied this procedure to the differently reduced data cubes mentioned in section~\ref{sec:observations}. We find that the differences in the results from these data cubes could be used to derive reasonable error estimates. We measure the position of the center to within 0.9\arcsec\ in the horizontal direction and 1.2\arcsec\ in the vertical direction. The position of the dynamical center coinsides with the dynamical nucleus found by Thatte et al. (2000). The average PA $= 219\pm 15\deg$ and average \Vsys\ $= 520\pm 14$ \kms, are fully consistent with the comprehensive \CO\ and \HI\ analysis by { L04b} (see their table~2). The ``$\phi$--versus--\Vlos'' plots (Krajnovi\'c et al. 2006) for each of the tilted rings clearly demonstrate the non-even distribution of the pixels at radii larger than 50\arcsec. Confirmed by the larger error bars, this implies comparably less reliable kinematic parameters at the outer radii. We mark these points in Fig.~\ref{fig:vrot}, and omit the outer regions in our angular frequency study in Fig.~\ref{fig:op}. \looseness-2

The derived kinematic paramaters shown in Fig.~\ref{fig:vrot} display some prominent features which indicate significant mass transfer in the disk of M\,83. These are smooth PA variation of $55\deg$, \Vsys\ variation of 40 \kms, and \Vrad\ variation of $60$ \kms, in the central $20$\arcsec. All these features are accompanied by the prominent drop of the \Vrot\ at the same radius, in turn consistent with the derived outer ILR radius found in Fig~\ref{fig:op}. These features can also be seen in the two-dimensional velocity field as a spiral pattern superimposed on a rapidly rotating inner component. \looseness-2

\subsection{The Nuclear Disk}
\label{sec:disk}
The rapidly rotating nuclear component in M\,83 was first found in the \CO\ observations of {L04b} and Sakamoto et al. (2004) as an inner disk with an exponential disk scale length of 50 pc enclosing $3\times10^8 M_\sun$. The nuclear disk includes only $\approx 5\%$ of the total molecular and neutral gas mass of M\,83, which in turn is not more than 10\% of the total galaxy mass (Huchtmeier \& Bohnenstengel 1981; Crosthwaite et al. 2002; Lundgren et al. 2004a, hereafter {L04a}). The \ghafas\ data resolves the nuclear disk to an unprecedented level, the kinematic signature of which we quantify by fitting an exponential disk profile to the central region of the velocity field (Fathi et al. 2006). We fix the inclination throughout the galaxy disk and assume that the nuclear disk component has a mass-to-light ratio of unity and dominates the nuclear regions of the galaxy. The observed velocity information can then be used to derive the disk parameters (e.g., Fathi et al. 2005;2006), and the errors can be estimated by varying the size of the fitting region and using the different sky-subtractions and binning schemes described in section~\ref{sec:quantify}. We derive the nuclear disk kinematic major axis of $230\pm 11\deg$, systemic velocity of $513\pm 5$ \kms, scale length of $60 \pm 20$ pc, and disk mass of $5.5 \pm 0.9 \times 10^8 M_\sun$. For a homogenously increased mass-to-light ratio by a factor two in the nuclear 60 pc of M\,83, the derived dynamical disk mass decreases by a factor two and vice versa for the case of a mass-to-light ratio smaler than unity. We see no evidence for the presence of dark matter in the nucleus of this galaxy, however, the nuclear starburst population could lead to a mass-to-light ratio less that one.\looseness-2

\subsection{The Spiral Inflow}
\label{sec:inflow}
\begin{figure}
\includegraphics[width=0.49\textwidth]{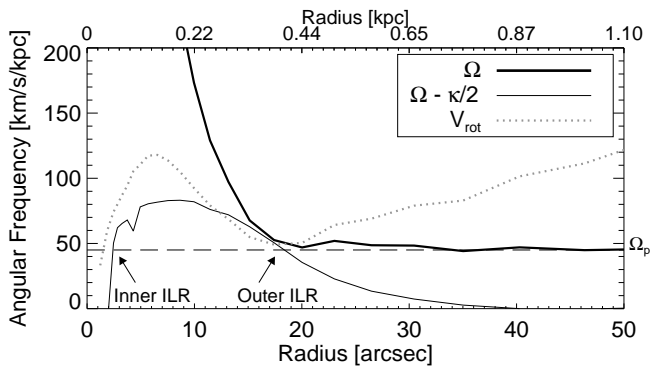}
\caption{Zooming into the central kpc of the angular frequency curve together with the effect of the epicyclic frequencies.}
\label{fig:op}
\end{figure}

In M\,83, spiral inflow on galactic scales has been reported by {L04b}. After removing the nuclear disk, our observed velocity field displays a spiral shape of the \Ha-emitting gas, perfectly aligned with the large-scale spiral inflow displayed in \CO\ (Fig.~8 of {L04b}). The unprecedented resolution of \ghafas\  (9 pc/pix at the galaxy, c.f., $\approx 500$ pc/pix of the \CO\ data) allows us to follow the inflow along a spiral trajectory down to a few tens of pc from the dynamical nucleus of M\,83. If the smooth kinematic PA twist in the central $20\arcsec \approx 400$ pc is associated with a trailing spiral pattern in the circumnuclear regions of M\,83 (aligned with the large scale spiral structure of M\,83), the streaming motions indicate gas inflow along spiral paths (e.g., Wong et al. 2004; Maciejewski 2004; Yuan \& Yang 2006). The harmonic decomposition allows us to measure a \Vrad\ as high as 50 \kms\ within the same region. The amplitude of the streaming motions is comparable to previously measured values on large scales as well as nuclear regions of spiral galaxies (e.g., Zurita et al. 2004; Fathi et al. 2005;2006). We use the \Vrot\ curve presented in Fig.~\ref{fig:vrot} together with the spiral structure pattern speed derived by Zimmer et al. (2004) to study the resonant interaction in the disk of M\,83. Fig.~\ref{fig:op} illustrates that for a pattern speed, $\Omega_\mathrm{p} = 45 \pm 8$ \kmskpc, our resolved rotation curve gives the position of an outer ILR radius at $\approx 20$\arcsec\ and an inner ILR radius at $\approx 2$\arcsec. Although slightly smaller, the outer ILR radius agrees well with the value estimated by {L04b}. We attribute the difference to beam smearing in the \CO\ data.\looseness-2

\section{Discussion and Conclusion}
\label{sec:discussion}
Classified as a starburst galaxy, M\,83 has all the ingredients to allow efficient mass transfer from kpc scale down to the nuclear region. The bar is the main actor in this scenario, driving the outer spiral arms as well as the inflow from its ILR toward the inner ILR radius. The highly obscured starburst activity in the central 50 pc radius has been forming stars at a rate of $\approx 3 \times 10^{-6} $ \mpcyr\  for at least $10^7$ years with signatures of prolonged star formation (Harris et al. 2001; Muraoka et al. 2007). This is a factor 30 higher than the star formation in the outer disk. The disk contains a large amount of atomic and molecular gas with significant molecular gas content in the stellar bar and in the nuclear region and a deficit of atomic gas in the central region (Sakamoto et al. 2004). The \CO\ kinematic analysis of {L04b} clearly showed that molecular gas spirals in from the outer regions down to a galactocentric distance of $\approx 200$ pc, confirming that gas can be transported efficiently from kpc scales all the way down to hundreds of pc scales close to the nucleus. Our high resolution data allows us to follow this inflow along the spirals down to the central few tens of pc scales.\looseness-2

In case of the bar and spiral arms having the same pattern speed, supported by the absence of multiple pattern speeds (Zimmer et al. 2004), the bar in M\,83 could drive the outer spiral arms (Elmegreen \& Elmegreen 1995). Our study suggests that the bar also drives the spiral arms in its interior as well as the nuclear starburst. At the outer ILR, just before 20\arcsec, the well known patchy ILR ring of M\,83 is found (Elmegreen et al. 1998). Our derived inner ILR radius agrees with the location of the nuclear starburst described, among many others, by Harris et al. (2001) and Calzetti et al. (2004). The derived inner ILR radius agrees with the scale length of the inner disk (section~\ref{sec:disk}), confirming the build-up of the inner disk from the outer disk material by resonant interaction exerted by the main bar in M\,83.\looseness-2

We confirm the derived disk parameters by applying the same fit to the one-dimensional ``true'' rotation curve presented in Fig.~\ref{fig:vrot}, showing that the nuclear disk indeed dominates the nuclear regions of the velocity field. The PA and \Vsys\ of the nuclear disk confirm that it shares the same orientation as the outer disk, and most likely it also lies in the same plane, since it would have to be inclined by more than $75\deg$ to match the observed outer velocities in its vicinity. Its mass is consistent with the values given by {L04b} and Sakamoto et al. (2004), and it is centered on the dynamical nucleus of the galaxy.\looseness-2

The position and nature of the nucleus of M83 is somewhat controversial. Whereas the strongest optical/near-IR peak has traditionally been seen as the nucleus, and Thatte et al. (2000) were the first to suggest that M\,83 has a double nucleus, we recently postulated (Sharp et al. 2007) that in fact Thatte et al.'s "second nucleus" is the true, though rather heavily extincted, nucleus of M\,83, whereas the optical peak is a massive super star cluster. Our \Ha\ kinematics confirm the presence of the same dynamical center for both the outer and the inner disk. We also confirm the double-peaked dispersion profile suggested by Thatte et al. (2000), however, due to strong contamination by dust in our maps, we cannot delve into the morphology of the nucleus. Fathi et al. (2005) and Falc\'on-Barroso et al. (2006) showed that the presence of dust has a much smaller effect on the derived line-of-sight-velocities, as compared with morphology, ratifying the advantage of using the kinematics for deriving the position of the dynamical center. Furthermore, we observe a patchiness in the central parts of the residual velocity field (Fig.~\ref{fig:kinematics}), correlated with the presence of dust (Harris et al. 2001).\looseness-2

In a direct comparison with near-infrared observations, we (Sharp et al. 2007) find a patchy star forming arc clearly associated with dust 100 pc west of the dynamical nucleus (see section~\ref{sec:results}). The zero-velocity curve goes through the dynamical nucleus, twists, and then passes through the bright knot suggested as the optical nucleus at 3\arcsec\ north-east of it. These kinematic features show that the nuclear region of M\,83, interior to the inner ILR radius of the bar, is dynamically fed by gas from the main galaxy disk, containing multiple stellar populations which could be explained by a multiple-burst scenario (e.g., Sarzi et al. 2007). The age gradient across the starburst region, shown in Ryder et al. (2005) confirms such a scenario as a plausible one for M\,83, thus the nuclear star forming region hosts several bursts caused by varying inflow rate of the disk material onto the starburst.\looseness-2

\begin{acknowledgements}
We thank the anonymous referee for insightful comments which helped us to clarify the manuscript. It is a pleasure to thank the ING staff, in particular R. Corradi, J. Rey, and A. Cardwell for their support on La Palma. KF acknowledges support from the Wenner-Gren foundations. KF and JEB acknowledge financial support from the Spanish Ministry of Education and Science, and the IAC project P3/86.
\end{acknowledgements}

%%%%%%%%%%%%%%%%%%%%%%%%%%%%%%%%
%\usepackage{natbib}
%\bibpunct{(}{)}{;}{a}{}{,}
%%%%%%%%%%%%%%%%%%%%%%%%%%%%%%%%

%\bibliographystyle{aa}
%\bibliography{M83_Fathi_v0.0}

\end{document}